\documentclass[fleqn,twoside]{article}
\usepackage{espcrc2}

\usepackage{graphicx}
\usepackage[figuresright]{rotating}

\newcommand{\AmS}{{\protect\the\textfont2
  A\kern-.1667em\lower.5ex\hbox{M}\kern-.125emS}}

\title{Generating cosmological perturbations with mass variations}

\author{Filippo Vernizzi\address[MCSD]{Helsinki Institute of Physics, 
        P.O.\ Box 64,
FIN-00014 University of Helsinki - Finland}}

\newcommand{\g}{\gamma}

\newcommand{\be}{\begin{equation}}
\newcommand{\ee}{\end{equation}}
\newcommand{\gsim}{\raise.3ex\hbox{$>$\kern-.75em\lower1ex\hbox{$\sim$}}}
\newcommand{\lsim}{\raise.3ex\hbox{$<$\kern-.75em\lower1ex\hbox{$\sim$}}}
\newcommand{\ba}{\begin{eqnarray}}
\newcommand{\ea}{\end{eqnarray}}
\newcommand{\bea}{\begin{eqnarray}}
\newcommand{\eea}{\end{eqnarray}}
\newcommand{\bean}{\begin{eqnarray*}}
\newcommand{\eean}{\end{eqnarray*}}

\def\spose#1{\hbox to 0pt{#1\hss}}
\def\ltapprox{\mathrel{\spose{\lower 3pt\hbox{$\mathchar"218$}}
\raise 2.0pt\hbox{$\mathchar"13C$}}}
\def\gtapprox{\mathrel{\spose{\lower 3pt\hbox{$\mathchar"218$}}
\raise 2.0pt\hbox{$\mathchar"13E$}}}

\def\Tdot#1{{{#1}^{\hbox{.}}}}

\begin{document}

\begin{abstract}
We study the possibility that large scale cosmological
perturbations have been generated during the domination and decay
of a massive particle species whose mass depends on the
expectation value of a light scalar field. We discuss  the
constraints that must be imposed on the field in order to remain
light and on the annihilation cross section and decay rate of the
massive particles in order for the mechanism to be efficient. We
compute the resulting curvature perturbations after the mass
domination, recovering the results of Dvali, Gruzinov, and
Zaldarriaga in the limit of total domination. By comparing the
amplitude of perturbations generated by the mass
domination to those originally present from inflation,
we conclude that this mechanism can be the primary source of perturbations
only if inflation
does not rely on slow-roll conditions.

\end{abstract}

\maketitle

\section{Introduction}

Although inflation has recently become the paradigm of the early
universe and the dominant contender for generating the observed
large scale density perturbations \cite{Lyth}, it is known to be
plagued by the
difficulty of being embedded in most current particle physics
models.

The usual hypothesis is that primordial density perturbations
originate from the vacuum quantum fluctuations of the inflaton
field. The fact that the inflaton has to remain {\em light}
during inflation and that its vacuum energy is constrained by the
amplitude of the observed cosmological perturbations make it
difficult to construct motivated models \cite{Dimopoulos}. For
this reason,  scenarios where inflation does not
rely on slow-roll conditions have recently been proposed (see e.g.,
\cite{Dvali}).

If, as in these scenarios, the inflaton has no fluctuations, it is
necessary to find a mechanism to generates cosmological
perturbations. In the curvaton scenario
\cite{curvaton,curvatonplus} curvature perturbations are generated
by the late decay of a light scalar field $\phi$, whose
fluctuation $\delta \phi$ is amplified during inflation with a
quasi-scale invariant spectrum, and amplitude set by the Hubble
rate at horizon crossing, $\delta \phi \sim H$. This mechanism
opens up new possibilities for observations and for model
building. In particular, it can {\em liberate} the inflaton from
the requirement that it is responsible for the primordial
perturbations.

In the same spirit,  inspired by the idea that coupling constants
and masses of particles during  the early universe may depend on
the value of some light scalar field, a number of people have
proposed that  perturbations could be generated from the
fluctuations of the inflaton coupling to ordinary matter during
reheating, a mechanism called inhomogeneous reheating or modulated
fluctuations \cite{gamma} (see also \cite{JP,Tutti} and
\cite{Mazum,Tsuji} for the perturbations).

Here we study the possibility that cosmological perturbations are
generated during the phase of domination of a massive particle
species $\psi$ whose mass is modulated by the value of a light
field $\phi$, as originally proposed in \cite{gamma3}. These
particles are produced and thermalize after inflation. If
sufficiently long-lived, they can freeze out and eventually
dominate the universe when they become non-relativistic. Due to
the mass fluctuation, the mass domination process becomes
inhomogeneous and converts the fluctuation of $\phi$
into curvature perturbations.

In \cite{Mio} we studied the perturbations generated by this
mechanism making use of the formalism introduced in \cite{wmu} for
interacting fluids. Here we take a different approach and, in
particular, we consider more appropriate initial conditions for
the perturbations of the massive particles after inflation. Due to
this different choice of initial conditions, the result found here
for the final curvature perturbation in terms of the mass fluctuations
does not agree with \cite{Mio}, but confirms \cite{gamma3}.

In the next section we model the inhomogeneous mass domination.
We derive and discuss the conditions
that must be imposed on the field
for remaining light
(Sec.~\ref{lightfield})
and on the annihilation cross section and decay rate of
the massive particles in order for $\psi$ to
dominate the universe and for the mechanism to be efficient
(Sec.~\ref{sec:domination}).
Finally, we study the perturbations and the observational
consequences of this mechanism, and we discuss when it can
successfully liberate inflation.

\section{Modeling the mass fluctuation}

In this section we discuss the homogeneous evolution
equations governing a fluid of non-relativistic particles $\psi$
with a rest mass that depends on the value of a scalar field.
Since we want to describe the universe after reheating,
we also add a radiation fluid and let
$\psi$ to decay into radiation.
The derivation  given here of the coupling between $\psi$ and
$\phi$ is `intuitive': A more rigorous treatment can be found in
the literature on scalar-tensor theories such as in \cite{Damour,Peebles}.

We consider a flat Friedmann-Lema\^{\i}tre-Robertson-Walker 
universe, with metric
$ds^2 = -dt^2+a^2(t) dx^2$. The Hubble rate $H=\dot a /a$
($\Tdot{}\!= \partial / \partial t$) is given by
\be
\label{eq:Friedmann} H^2 = \frac{1}{3m^{2}_{\rm P}} (\rho_\psi +
\rho_\gamma + \rho_\phi ),
\ee
where $\rho_X$ is the energy density of the $X$ species and
$m_{\rm P}=(8 \pi G)^{-1/2}$ is the reduced Planck mass.

The non-relativistic particles considered here are similar to cold
dark matter particles: they are collisionless and
non-relativistic. Hence their pressure vanishes
and their energy density is given by the product of the
particle number density and the mass,
\be
T_{\mu \nu}^{(\psi)} =\rho_\psi u_\mu u_\nu = m_\psi n_\psi u_\mu u_\nu,
\label{eq:emdef}
\ee
where $u_\nu$ is the four-velocity. The mass of $\psi$
depends on a scalar field $\phi$, $m_\psi=m_\psi(\phi)$.

We consider $\psi$ after freeze-out so that $n_\psi$
is not changed by annihilation processes. However,
since we want to allow $\psi$ to decay into radiation, we define a
decay rate of change of the number of $\psi$ particles as $
\nabla_\mu (n_\psi u^\mu) = -\Gamma n_\psi $. With this equation,
 Eq.~(\ref{eq:emdef}) yields a conservation equation for the
energy of $\psi$, written in covariant form \cite{Mio},
\be
u^\nu \nabla_\mu T_{ \nu}^{(\psi)
\mu} =  - ( \alpha \phi^{-1}\partial_\mu \phi u^\mu - \Gamma)
\rho_\psi , \label{eq:1}
\ee
where we have defined the (dimensionless) mass coupling function
$
\alpha = {d \ln
m_\psi}/{d \ln \phi}$.
Even in the absence of coupling to radiation,
due to the interaction with $\phi$, the energy of $\psi$ is not
conserved.

Since $\psi$ decays only into radiation, the evolution equation
governing the radiation energy density is
\be
u^\nu \nabla_\mu T_{ \nu}^{(\gamma) \mu} =  -\Gamma \rho_\psi .
\label{eq:3}
\ee
The background conservation equations for the two fluids then
become (with $P_\gamma=\rho_\gamma/3 $)
\bea
\label{eq:dyn1}
\dot{\rho}_\psi &=& - (3H - \alpha \dot \phi/\phi + \Gamma)  \rho_\psi , \\
\label{eq:dyn2} \dot{\rho}_\gamma &=& - 4 H \rho_\gamma + \Gamma
\rho_\psi.
\eea

The scalar field $\phi$ has a standard kinetic term and a
potential $V(\phi)$. However, its evolution equation is sourced by
an additional term due to the coupling to the bath of
non-relativistic particles $\psi$. Indeed, the requirement that
the sum of the three energy-momentum tensors is conserved, $
 \sum_X \nabla_\mu T_{ \nu}^{(X) \mu}=0$, yields
\be
\nabla_\mu \partial^\mu \phi = V'+ (\alpha/\phi) \rho_\psi.
\label{eq:5}
\ee
The background Klein-Gordon equation for the scalar field in
discussed in the following section.

\section{Scalar field behavior}
\label{lightfield}

The background part of Eq.~(\ref{eq:5}) can be rewritten as
\be
\ddot \phi + 3H\dot \phi + V' = - (\alpha/\phi)
\rho_\psi,\label{eq:dyn4}
\ee
where ${}'\!= d/ d \phi$. Taking into account the
coupling to $\psi$, the scalar field behaves as a field in the
effective potential $V_{\rm eff} (\phi)=V(\phi) + \rho_\psi$. The
coupling can have the effect of increasing the mass of $\phi$ and
drive its evolution.

Exploring the behavior of the scalar field during the phase that
goes from the radiation dominated era to the mass domination
requires a form for $V(\phi)$ and $m_\psi(\phi)$. 
Several form of these functions have been explored in the literature.
If the effective mass of $\phi$ is made larger than $H$ by the 
coupling to $\psi$,
the field rapidly sets to the
minimum of $V_{\rm eff}$ \cite{Carroll,Khoury}. 
Here, however, we want to protect the mass of $\phi$ and maintain
the field light, to conserve the primordial 
fluctuations inherited from inflation \cite{gamma3}. 
This puts strong
constraints on the initial value of $\phi$.

We consider a massive field, $V(\phi)=\frac{1}{2}m_\phi^2 \phi^2$,
with $m_\phi < \Gamma$, so that $m_\phi < H$ until the decay of
$\psi$. The condition for the field to remain light is then
\be
{\phi}/{ m_{\rm P}} > (\Omega^{\rm dec}_\psi |\alpha|)^{1/2},
\label{eq:remainlightzero}
\ee
where $\Omega^{\rm dec}_\psi$ is the abundance of $\psi$, 
$\Omega_\psi=\rho_\psi/\rho$, at its decay. A more tighter condition is to
impose that $V'$ always dominates over $(\alpha/\phi) \rho_\psi$,
\be
{\phi_{\rm in}}/{ m_{\rm P}} > (\Omega^{\rm in}_\psi |\alpha|)^{1/2}
H_{\rm in}/m_\phi. \label{eq:remainlight} 
\ee
If this condition holds
at some time `in' 
deep in the radiation era, it is maintained throughout the
domination.  
These conditions can be realized for $\phi \sim m_{\rm P}$ if
$\Omega_\psi \alpha$ is small enough. 
An
analogous condition on $\phi$ was first advocated in \cite{gamma3}
in order to protect its mass from acquiring too large thermal
corrections due to the coupling with the bath of $\psi$ particles.

If, on the other hand, $\phi$ is too large, the scalar field can
dominate the universe by entering a second inflationary era. We
must hence require that
\be
{\phi}/{ m_{\rm P}} < {\Gamma}/{m_\phi}.
\ee

In the following we assume the stronger condition
(\ref{eq:remainlight}). This simplifies considerably the treatment
of perturbations than by only requiring the weaker condition,
Eq.~(\ref{eq:remainlightzero}). One can indeed study the background
evolution of $\phi$ neglecting the coupling to $\psi$.
Equation~(\ref{eq:dyn4}) becomes
\be
\ddot \phi + 3{\beta}\dot \phi /{t}   +m_\phi^2 \phi \simeq 0,
\quad a(t) \propto t^{\beta}. \label{eq:light}
\ee
The solution of this equation (regular for $m_\phi t \to 0$) is
given in terms of the Bessel function $J_\nu$ with $\nu=(3
\beta-1)/2$: $\phi \simeq \phi_* A_\nu J_{\nu}(m_\phi t)/(m_\phi
t)^{\nu}$, where $A_\nu$ is a numerical constant and the star
stands for the value during inflation. Hence we obtain
\be
3H \dot \phi \simeq -\frac{3\beta}{3\beta+1} m_\phi^2 \phi,
\quad m_\phi t\ll 1. \label{shown}
\ee
Note that this solution is not of `slow-roll type': The
acceleration $\ddot \phi$ is not small with respect to $H \dot \phi $. 
However, as
during slow-roll, the kinetic energy of the scalar field is
subdominant with respect to the potential energy, $ P_\phi
 \simeq - \rho_\phi$, while the adiabatic speed of sound
depends on $\beta$, $c_\phi^2\simeq-1-2/(3 \beta)$.

\section{Conditions for the mass domination}
\label{sec:domination}

For the mass fluctuations to be imprinted into the density perturbations
it is essential that $\psi$ dominates or at least becomes a significant
component of the energy density of the universe before its decay.
If $\psi$ froze out at a temperature $T$ such that $m_\psi/T$ was
not much larger than 1, then the species can have a significant
relic abundance and, if sufficiently long-lived, can eventually
dominate the universe before decaying. Here we want to study when
this condition is realized.

As shown by Eq.~(\ref{shown}), the evolution of the light field
is slow with respect to the expansion rate, $-\dot \phi/\phi \sim
(m_\phi/H)^2 H \ll H$, so that the mass variation is also slow,
$\dot m_\psi/m_\psi\ll H$. Furthermore, we are considering
non-relativistic particles, i.e., $T<m_\psi \Rightarrow H <
m_\psi^2/m_{\rm P} \ll m_\psi$. Thus, when studying their
abundance, the particles $\psi$ can be
considered as having constant mass 
\cite{Carroll}, and we can consistently neglect the
field coupling on the right hand side of Eq.~(\ref{eq:dyn1}).

\begin{figure}
\begin{center}
\includegraphics*[width=7.5cm]{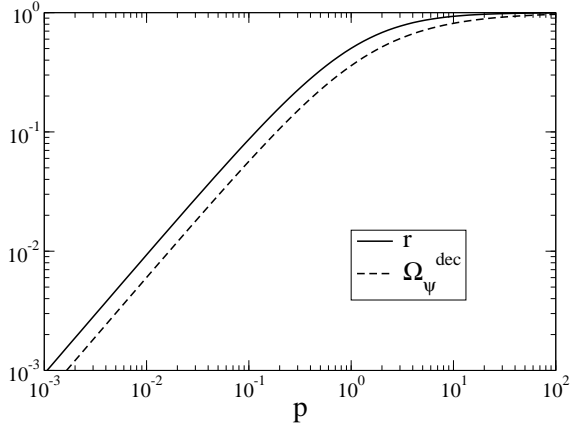}
\caption{The dashed line shows the abundance of $\psi$ at its decay, 
$\Omega_\psi^{\rm
dec}$, as a function of the initial
parameter $p$. This is compared to the efficiency parameter $r$
(solid line), defined in
Sec.~\ref{numeri}.} \label{fig:r}
\end{center}
\end{figure}

The system of equations (\ref{eq:dyn1}) and (\ref{eq:dyn2}) then 
describes the decay of a dust
fluid into radiation and can be easily solved numerically (see e.g.,
\cite{wmu,Mio}). The initial conditions for this system are taken well into 
the radiation era, $\Omega_\psi \ll 1$,
once $\psi$ has frozen out. 
As shown in Ref.~\cite{wmu}, close to the initial condition
$\Omega_\psi \ll 1$, $\Omega_\psi \simeq p (\Gamma/H)^{1/2}$.
The value of the parameter
\be
p = \Omega^{\rm in}_\psi ({H_{\rm in}}/{\Gamma})^{1/2}
, \label{eq:Rin}
\ee
determines the trajectories followed by the system on the
$(\Omega_\psi, H)$-plane. In particular it determines the abundance of $\psi$ 
at its decay, $\Omega_\psi^{\rm dec}$. This is shown in Fig.~\ref{fig:r} as
a function of $p$. If initially $\Gamma \gg H \Omega_\psi^2$,
i.e., $p\ll 1$, then the decay is almost instantaneous and $\psi$
does not have the time to dominate, $\Omega_\psi^{\rm dec}\ll 1$.
On the contrary, when $p$ is initially large, then
$\Omega_\psi^{\rm dec}=1$.
Thus, the condition for $\psi$ to dominate the universe 
becomes $p \gtapprox 1$.

Now we want to translate this into a condition on the decay rate
and annihilation cross section of $\psi$. On using
the definition of $p$, Eq.~(\ref{eq:Rin}), we can write the
condition for the domination as
\be
p\simeq g^{1/4}  \frac{n_\psi}{s} \frac{m_\psi}{\sqrt{m_{\rm P}
\Gamma}} \gtapprox 1 ,\label{eq:kkk}
\ee
where $g$ is the number of relativistic degrees of freedom and
$n_\psi/s$ is the relic abundance of the $\psi$ particles when
they freeze-out -- $s$ is the entropy density. In order to derive
(\ref{eq:kkk}) we have used $\rho \simeq \rho_\g \simeq g T^4$,
and $s \simeq g T^3$.

After freeze-out and for $\Gamma \ll H$, the relic abundance
$n_\psi/s$ is constant (we assume $g$ constant). The relic
abundance can be evaluated by solving the Boltzmann equation
describing the freeze out of the annihilation of $\psi$ particles
and antiparticles with cross section $\sigma_A$, which we take
to be independent of the particles energy. In this case there is
an approximate solution for the abundance at freeze-out
\cite{JGK},
\be
\frac{n_\psi}{s} \simeq \frac{100}{m_\psi m_{\rm P} g^{1/2}
\langle \sigma_A v \rangle},
\ee
where $\langle \sigma_A v \rangle$ is the thermal average of the
total cross section times the relative velocity $v$. On using this
solution, Eq.~(\ref{eq:kkk}) becomes independent of the mass
$m_\psi$ \cite{Mio},
\be
p \simeq \frac{100}{g^{1/4}} \left( \frac{ m_{\rm P}}{\Gamma}
\right)^{1/2} \frac{ m_{\rm P}^{-2} }{\langle \sigma_A v \rangle }
\gtapprox 1.
\ee
This relation holds if the $\psi$-particles are subdominant at
freeze-out. The smaller the annihilation cross section is, the
earlier $\psi$ freeze out before becoming non-relativistic and the
larger is $\Omega^{\rm dec}_\psi$.

On requiring that the massive particles decay before
nucleosynthesis, i.e.,  $\Gamma \gtapprox (1 {\rm MeV})^2/m_{\rm
P}$, we find that only for $\langle \sigma_A v \rangle \ltapprox
10/g^{1/2} (10^7 {\rm GeV})^{-2}$ does $\psi$ dominate the
universe. Hence, the initial thermal equilibrium by annihilation
of particles and antiparticles must be maintained by some gauge
interaction which is weaker than those of the standard model. This
excludes that $\psi$ is made of standard model particles.

\section{Cosmological perturbations from the mass domination}

In this section we derive the coupled perturbed equations of
motion for the two fluids and the scalar field. Then we
discuss their solutions, analytically and numerically.
The aim is to study the mechanism of conversion of scalar field
fluctuations into curvature perturbations during the mass
domination.

We describe scalar perturbations in the metric with
line element
\begin{equation}
ds^2=-(1+2\Phi)dt^2
+a^2 (1-2\Phi)\delta_{ij} dx^idx^j ,
\end{equation}
which follows from the absence of anisotropic stress
perturbations. The quantity $\Phi$ corresponds to the Bardeen
potential in longitudinal gauge.

In order to perturb the energy-momentum tensor of the three
components $\psi$, $\g$, and $\phi$, we introduce the energy
density and pressure perturbations,  $\delta \rho_X$ and $\delta
P_X$, and the scalar field perturbation $\delta \phi$. It is also
useful to define the relative perturbation $\delta_X=\delta
\rho_X/\rho_X$. In terms of these quantities, we perturb
Eqs.~(\ref{eq:1}), (\ref{eq:3}), and (\ref{eq:5}), and since we
are interested in the large scale perturbations,  we drop the
gradient terms, that vanish in the large scale limit. This yields
\bea
\dot \delta_\psi - 3 \dot \Phi=
\Tdot{\left(\alpha \frac{\delta \phi}{\phi} \right)} -\Gamma \Phi ,
\label{eq:first}\\
\dot \delta_\g - 4 \dot \Phi= \Gamma \frac{\rho_\psi}{\rho_\g}
(\delta_\psi - \delta_\g +\Phi), \label{eq:second} \\
\ddot {\delta \phi} + 3 H \dot {\delta \phi} +
\left(m_\phi^2 + \frac{\Tdot{ (\alpha/\phi)}}{\dot \phi} \rho_\psi \right) \delta \phi= \quad
\nonumber\\
4 \dot \Phi \dot \phi -2 V' \Phi - (\alpha/\phi)
\rho_\psi(\delta_\psi + 2 \Phi) .\label{eq:third}
\eea
The perturbed energy constraint equation reads
\be
6H(\dot \Phi +H\Phi) =-m^{-2}_{\rm P}(\rho_\psi \delta_\psi
+\rho_\g \delta_\g+\rho_\phi \delta_\phi). \label{eq:last}
\ee

Equation~(\ref{eq:third}) can be opportunely simplified: Since the
scalar field is light and subdominant we can neglect 
the right hand side. Assuming the strong constraint 
(\ref{eq:remainlight})
we can also neglect the mass correction due to the coupling.
Equation~(\ref{eq:third}) then simplifies to the equation of a
light test field,
\be
\ddot {\delta \phi} + 3 H \dot {\delta \phi} + m_\phi^2  \delta
\phi = 0,\label{eq:simple3}
\ee
that in the massive case is the same as the equation for the
background field. Hence, the ratio between
$\delta \phi$ and $\phi$ is constant in time,
\be
\delta \phi/\phi = \delta \phi_*/\phi_*. \label{eq:consratio}
\ee
Assuming only the weaker constraint 
(\ref{eq:remainlightzero}) implies that 
$\delta \phi/\phi$ may slowly vary, which makes the mechanism
more difficult to study, although it maintains its qualitative features.

\subsection{Analytic calculation}

It is possible to analytically solve the system of equations
(\ref{eq:first}), (\ref{eq:second}), and (\ref{eq:last}), in the
case that the massive particles dominate completely the universe
before decaying. We can hence consistently assume that we are far
from the decay, $H\gg \Gamma$, so that the source terms
proportional to $\Gamma$ in Eqs.~(\ref{eq:first}) and
(\ref{eq:second}) can be neglected for the moment.
Equations~(\ref{eq:first}) and (\ref{eq:second}) then simplify,
\bea
\dot \delta_\psi - 3 \dot \Phi &=&
\dot \alpha {\delta \phi_*}/\phi_*  , \label{eq:simple1}\\
\dot \delta_\g - 4 \dot \Phi&=& 0, \label{eq:simple2}
\eea
where we have used 
Eq.~(\ref{eq:consratio}).

We first discuss the {\em initial conditions} of these equations.
They are defined in the radiation dominated era, just after the
freeze-out of $\psi$. In this limit the Bardeen potential $\Phi$
is constant, and the perturbation of the radiation fluid can be
found from the constraint (\ref{eq:last}), on using $\rho \simeq
\rho_\g$. This yields $\delta^{(i)}_\g = -2 \Phi_*$, where $\Phi_*$
is the value of $\Phi$ at the beginning of the radiation dominated
era (e.g., after inflation). This initial perturbation is
usually dropped and considered negligible in the treatment of
models where perturbations are produced after inflation by a light
field. Here however, in the spirit of \cite{curvadavid}, we retain
this term since it can turn out to be larger than the curvature
perturbation generated by the mass domination (see below).

The value of the initial condition $\delta_\psi^{(i)}$ is crucial
to correctly determine the final curvature perturbation produced.
Unfortunately it is not unambiguously defined. Here we set $\delta_\psi^{(i)}$
by assuming
that radiation and $\psi$ particles were created by the same
mechanism (e.g., reheating) with the {\em same number density
perturbation},\footnote{In Ref.~\cite{Mio} the initial condition
$\delta_\psi/3 = \delta_\g/4$, i.e., absence of relative entropy
perturbation between $\psi$ and $\g$, was used. Only if $m_\psi$
is constant, is this condition the same as
Eq.~(\ref{eq:conditionE}). The choice of Eq.~(\ref{eq:conditionE})
is  at the origin of the discrepancy between the result of this
work and the result of \cite{Mio} on the final total curvature
perturbation.}
\be
\delta \left(\frac{n^{(i)}_\psi}{n^{(i)}_\g} \right)=0 
\quad \Rightarrow \quad \frac{\delta
n^{(i)}_\psi}{n^{(i)}_\psi} = \frac{3}{4} \delta^{(i)}_\g.
\label{eq:conditionE}
\ee
Thus, considering also the mass fluctuation, the relative initial
density perturbation of $\psi$ reads
\be
\delta^{(i)}_\psi = \frac{\delta m^{(i)}_\psi}{ m^{(i)}_\psi} +
\frac{\delta n^{(i)}_\psi }{n^{(i)}_\psi} = \alpha_* \frac{\delta
\phi_*}{\phi_*}+ \frac{3}{4} \delta^{(i)}_\g.
\label{eq:conditionA}
\ee
The initial conditions are thus given by
\bea
\Phi^{(i)}
=\Phi_*, \quad \delta \phi^{(i)}
 = \delta \phi_*,
\quad \dot {\delta \phi}^{(i)} =0, \nonumber\\
\delta_\psi^{(i)} = \alpha_* \delta \phi_*/\phi_* - (3/2) \Phi_*, \quad
\delta_\g^{(i)} = -2 \Phi_*.
\eea

Equations (\ref{eq:simple1}) and (\ref{eq:simple2}) can be easily
solved with these initial conditions, yielding
\bea
\delta_\psi &= &
\alpha \frac{\delta \phi_*}{\phi_*}
+3 \Phi- \frac{9}{2}\Phi_*, \label{eq:filo1} \\
\delta_\g &=& 4 \Phi - 6 \Phi_* ,\label{eq:filo2}
\eea
where the terms proportional to $\Phi_*$ are integration
constants. Only the first of these equations is used in the
following. At the final stage of the mass domination, when the
non-relativistic particles dominate the universe, the constraint
(\ref{eq:last}) yields $\delta_\psi = -2 \Phi$. Combined with
Eq.~(\ref{eq:filo1}) it implies that the final curvature
perturbation is
\be
\Phi = \frac{9}{10}\Phi_* - \frac{1}{5} \alpha
\frac{\delta \phi_*}{\phi_*} . \label{eq:result}
\ee
This result holds during the matter ($\psi$) dominated era, before
the decay of the massive particles. It is related to the curvature
perturbation in the subsequent radiation era after
$\psi$ decay by the well-known relation $\Phi_r = (10/9) \Phi_m$
\cite{Lyth}.

Equation (\ref{eq:result}) can be expressed in terms of the
curvature perturbation defined on the uniform density hypersurface,
$\zeta= - \Phi + {\delta \rho}/[{3(\rho + P)}]$ \cite{Bardeen},
which is known to be conserved for adiabatic perturbations. In the
radiation dominated era $\zeta=-(3/2) \Phi_r$, which yields
\be
\zeta= \zeta_* + \frac{1}{3} \left.\frac{\delta
m_\psi}{m_\psi}\right|_{\rm dec}  \quad (\Omega_\psi^{\rm
dec}=1). \label{eq:finaldom}
\ee
The coefficient $1/3$ in this equation confirms the result
of \cite{gamma3}.

\subsection{Numerical results}
\label{numeri}

Here we solve Eqs.~(\ref{eq:first}), (\ref{eq:second}), and
(\ref{eq:last}) numerically, so that we can study 
the case where the massive
particles do not dominate the universe completely  before
decaying. We initially consider the particular case of $\alpha=$ constant
(e.g., $m_\psi = \lambda \phi^n$). In this case the first term on
the right hand side of Eq.~(\ref{eq:first}) vanishes. The problem
then reduces to the study of a pressureless fluid with
non-vanishing initial perturbation (\ref{eq:conditionA}) that
decays into a relativistic fluid $\g$.

This problem has been studied in \cite{wmu} in the case of the
curvaton field. There, it was found that the final curvature
perturbation is $\zeta = \frac{1}{3}r \delta_\psi$ ($\Phi_*=0$ was
assumed), where $r$ is an efficiency parameter which can be
computed as a function of $p$. We have solved
Eqs.~(\ref{eq:first}) and (\ref{eq:second}) numerically and the
solution is
\be
\zeta=  \zeta_* +  \frac{r}{3} \alpha_* \frac{\delta
\phi_*}{\phi_*} , \quad (m_\psi=\lambda \phi, \ \ r \simeq
\Omega_\psi^{\rm dec}), \label{linearnondom}
\ee
with $r \le 1$ represented in Fig.~\ref{fig:r}. Here an initial
non-vanishing perturbation $\zeta_*$ has also been included.
Fig.~\ref{fig:r} shows that $r \simeq \Omega_\psi^{\rm dec}$. As
expected, the more $\psi$ dominates the universe before decaying,
the more the mechanism of generation of curvature perturbations is
efficient.

If $\alpha$ varies slowly during the phase of domination and
decaying, we can extend the results of the numerical
calculation to a general $\alpha$. We find
\be
\zeta= \zeta_* +
\frac{r}{3} \alpha_{\rm dec}  \frac{\delta \phi_*}{\phi_*}
= \zeta_* + \frac{r}{3} \left. \frac{\delta
m_\psi}{m_\psi}\right|_{\rm dec}
. \label{eq:final}
\ee
This is our main result. It generalizes the results of
\cite{gamma3} to the case where the non-relativistic particles do
not dominate the universe completely before decaying into
radiation.

\section{Non-Gaussianities}

Here we discuss an unambiguous observational consequence of the
mass domination, namely the presence of non-Gaussianities in the
perturbations. The non-Gaussianities generated by
the mass domination have been first computed in \cite{gamma3} in the
case of total domination ($\Omega_\psi^{\rm dec}=1$) (see also
\cite{Zalda,RiottononG}). The possible presence of isocurvature perturbations
has been studied in \cite{Mio}. 

For convenience we define $\tilde
r = r \alpha_{\rm dec} \simeq \Omega_\psi^{\rm dec} \alpha_{\rm
dec}$. 
We also define the ratio between the
perturbation generated by the mass domination and the primordial
curvature perturbation $\zeta_*$ [see Eq.~(\ref{eq:final})],
\be
R=\frac{1}{\zeta_*} \frac{\tilde r}{3} \frac{\delta
\phi_*}{\phi_*} . \label{eq:R}
\ee

We can parameterize the level of non-Gaussianities with the
non-linear parameter $f_{\rm NL}$ defined as $\zeta=\zeta_{\rm L} -({3}/{5})
f_{\rm NL} \zeta_{\rm L}^2$ \cite{KomaSper,Kamiol}, where $\zeta_{\rm L}$
represents the Gaussian (linear) contribution to the total
curvature perturbation $\zeta$. 
Motivated by inflation, we assume
$\zeta_*$ to be Gaussian.

Non-Gaussianities arise when the field perturbation $\delta
\phi_*$ becomes larger than $\sim 10^{-4} \phi_*$. This
happens when the efficiency is low, $\tilde r \ll 1$. In this case,
terms quadratic 
in $\delta \phi_*/\phi_*$, that have been neglected
in the linear calculation, become important
\cite{curvatonplus,gamma3},
\be
\zeta = \frac{\tilde r}{3}\left( \frac{\delta \phi_*}{\phi_*} +
\frac{1}{2} \frac{\delta {\phi_*}^2}{{\phi_*}^2} \right) +
\zeta_*.
\ee
This yields the non-linear parameter
\be
f_{\rm NL} = -\frac{5}{2\tilde r} (1+R^{-1})^{-2} .
\ee
As expected, when $R \to 0$, $f_{\rm NL} \to 0$. If, on the contrary,
$R$ is large, i.e., $\zeta$ is mainly due to the mass domination,
$f_{\rm NL} = -{5}/({2\tilde r})$ \cite{gamma3}. As noticed in
\cite{curvatonplus,gamma3} this number is in the ballpark of future
experiments like Planck, that will be able to detect $f_{\rm NL}
\gtapprox 5$ at $2\sigma$-level \cite{KomaSper}. Therefore we may
have a well detectable signature of non-Gaussianities if $\tilde r
\ltapprox 0.5$.

Here we are also interested in the lower bound 
on $\tilde r$ from the current bound on non-Gaussianities.
The WMAP limit on $f_{\rm NL}$ corresponds to $-58<
f_{\rm NL}<134$ ($2\sigma$-level) \cite{Komatsu}, which translates
into $\tilde r \gtapprox 0.02$.

\section{Liberating inflation?}

As mentioned in the introduction, the main reason to use a light
field other than the inflaton as a source of curvature
perturbations is to liberate the inflaton from the task of
generating the observed amplitude of perturbations.

In \cite{curvadavid} (see also \cite{mixed}) it was shown that, if
cosmological perturbations are produced during the late decay of a
light scalar field whose value during inflation is of the order of
$m_{\rm P}$, then the primordial curvature perturbation already
present from inflation, i.e., $\zeta_*$, is of the same order of
magnitude as the perturbation produced by the late decay. This is
true only if inflation is of slow-roll type, and
fluctuations in the inflaton field are $\sim H_*$, i.e., of the same
amplitude as the fluctuations in the light field $\phi$. It is
thus interesting to compare these two contributions in the context
of the mass domination.

The curvature perturbation $\zeta_*$ from inflation is, at first
order in the slow-roll parameters,
$\zeta_* = ({1}/{\sqrt{2 \epsilon}}) ({H_*}/{m_{\rm P}})$ \cite{Lyth},
where $\epsilon \ll 1$ is the first slow-roll parameter. We can replace
this expression in the definition of $R$, Eq.~(\ref{eq:R}),
\be
R= \tilde r \frac{\sqrt{2 \epsilon}}{3} \frac{m_{\rm P}}{\phi_*}. \label{eq:RR}
\ee
Unless $\phi_*$ is very small, $\phi_* \sim \tilde r
\sqrt{\epsilon} m_{\rm P}$, the perturbation generated by the mass
domination represents only a negligible correction to the
perturbation originally present from inflation. However, we know
that $\phi_*$ cannot be too small: 
Indeed, Eq.~(\ref{eq:remainlightzero}) implies that 
$ \phi_*/ m_{\rm P} > \tilde r^{1/2}$, 
which combined with Eq.~({\ref{eq:RR}}) yields
\be
R \ltapprox 0.5 \tilde r^{1/2}\sqrt{\epsilon} .
\ee
The stronger condition (\ref{eq:remainlight}) implies an even lower value of 
$R$.
We conclude that if inflation is of slow-roll type, the mass
domination mechanism does not provide a sufficient amplitude of
fluctuations to liberate inflation.

This conclusion however changes if  the inflaton does not go
through a standard slow-rolling phase, as proposed in
\cite{Dvali}. If the inflaton is not light during inflation, its
fluctuation is suppressed \cite{Locked} and the primordial
curvature perturbation originated during inflation is negligible,
$\zeta_* \simeq 0$. Only in this case, the inhomogeneous mass
domination mechanism represents a viable scenario for the
generation of primordial cosmological perturbations.

\section{Conclusion}

We have considered the possibility that cosmological
perturbations are generated by the mass domination
mechanism, as proposed in \cite{gamma3},
during a phase of domination
of non-relativistic particles whose mass is fluctuating in time and space,
modulated by a light scalar field.
By requiring that the scalar field remains light until the massive
particles decay and does not dominate the universe with a second
stage of inflation, we have shown that its value must be of the order
of the Planck mass. A further condition
must be imposed on the annihilation cross section 
of the massive particles. This must
be weak enough -- weaker than for standard model interactions --
as to let the massive particles freeze out
before being completely diluted when they become non-relativistic.

We show that if these conditions are met, a curvature perturbation
is produced. This is proportional to the mass fluctuation and
the abundance of massive
particles at their decay, as given by Eq.~(\ref{eq:final}), and 
confirms \cite{gamma3}, in the limit where the massive
particles completely dominate the universe.
Non-Gaussianities are inversely proportional to
the abundance of massive
particles at their decay, so that the latter can not be too small.

Perturbations produced during inflation add to those
generated by the mass domination.
If inflation if of slow-roll type,
the former are much larger than the latter. Only for inflationary
models that violate the slow-roll conditions is the
mass domination a successful mechanism for generating
the observed cosmological  perturbations.

\end{document}